\documentclass[%
 aip,
 amsmath,amssymb,
 preprint,%
]{revtex4-1}

\usepackage[dvipdfmx]{graphicx}
\usepackage{dcolumn}
\usepackage{bm}

\usepackage[utf8]{inputenc}
\usepackage[T1]{fontenc}
\usepackage{mathptmx}

\usepackage{ulem}
\usepackage{color}

\begin{document}

\preprint{AIP/123-QED}

\title{Thin Film Growth of Heavy Fermion Chiral Magnet YbNi$_3$Al$_9$}

\author{Hiroaki Shishido}
	\email{shishido@pe.osakafu-u.ac.jp}
	\affiliation{Department of Physics and Electronics, Graduate School of Engineering, Osaka Prefecture University, Sakai, Osaka 599-8531, Japan}
	\affiliation{NanoSquare Research Institute, Osaka Prefecture University, Sakai, Osaka 599-8531, Japan}
\author{Akira Okumura}
	\affiliation{Department of Physics and Electronics, Graduate School of Engineering, Osaka Prefecture University, Sakai, Osaka 599-8531, Japan}
\author{Tatsuya Saimyoji}
	\affiliation{Department of Physics and Electronics, Graduate School of Engineering, Osaka Prefecture University, Sakai, Osaka 599-8531, Japan}
\author{Shota Nakamura}
\author{Shigeo Ohara}
	\affiliation{Department of Physical Science and Engineering, Graduate School of Engineering, Nagoya Institute of Technology, Nagoya 466-8555, Japan} 
\author{Yoshihiko Togawa}
	\affiliation{Department of Physics and Electronics, Graduate School of Engineering, Osaka Prefecture University, Sakai, Osaka 599-8531, Japan}

\date{\today}

\begin{abstract}
We grew thin films of a heavy fermion chiral magnet YbNi$_3$Al$_9$ by using molecular beam epitaxy. 
They were grown on $c$-plane sapphire substrates under ultra-high vacuum while maintaining a deposition rate at a stoichiometric ratio among Yb, Ni, and Al.
The resulting thin films contain epitaxial grains with a $c$ axis parallel to the substrate surface: The YbNi$_3$Al$_9$ $c$ axis is parallel to the sapphire $b$ or $a$ axis. 
The temperature dependence of the resistivity exhibits a typical feature of a dense Kondo system with a broad shoulder structure at
$\sim$40\,K, as well as a kink as a signature of the chiral helimagnetic ordering at 3.6\,K.
These features are consistent with those previously observed in bulk samples.
The shift in the kink associated with the field-induced phase transition is found in the magnetoresistance curves under a magnetic field applied in the direction perpendicular to the $c$-axis. 
The magnetic phase diagram well reproduces that for the bulk crystals, implying that the chiral soliton lattice phase arises under magnetic fields, even in thin films. 
\end{abstract}

\maketitle
Chirality is a universal concept of symmetry found in a wide variety of materials, from molecules to crystals, and plays an important role in determining physical properties. 
It has attracted much attention in various research fields such as biology, pharmaceuticals, and optics. 
In the field of spin electronics, chiral magnetic materials exhibit nontrivial chiral magnetic ordering and substantial physical responses. 
Much effort has been devoted to the study of chiral magnetism from the viewpoints of fundamental physics and device applications.

Chiral helimagnetic (CHM) order appears as a ground state of chiral magnets at zero magnetic field below a magnetic ordering temperature, $T_m$. In the CHM state, the arrangement of neighboring magnetic moments is twisted in a fixed rotation sense along the screw axis of the crystal \cite{Tog16}. 
In the presence of magnetic fields, the CHM state exhibits nontrivial transformations into different kinds of chiral spin order, such as chiral conical phase, chiral soliton lattice (CSL), and chiral magnetic skyrmions. 
Interestingly, the enhancement of physical responses upon the formation of chiral spin order has been reported. 
For instance, nonreciprocal electrical conductivity with respect to the magnetic field and current directions, the so-called electrical magnetochiral effect \cite{Rik01, Aok19}, is found to be intensified in the chiral conical phase \cite{Aok19, Oku19}.

The formation of magnetic skyrmions, which are topological objects of chiral spin swirling \cite{Bog87, Bog94}, has been reported in cubic chiral crystals (e.g., MoGe \cite{Muh09, Yu10}, FeGe \cite{Yu12, McG16}, Fe$_{1-x}$Co$_x$Si \cite{Yu10_2}, Cu$_2$OSeO$_3$ \cite{Sek12, Ada12}, GaV$_4$S$_8$ \cite{Kez15}, and EuPtSi \cite{Kan19}) and ultra-thin films of transition metals \cite{Hei11}.
Demonstrating the particle-like nature of magnetic skyrmions, they can be created and annihilated \cite{Sam13, Oka16} as well as driven \cite{Jon10, Sam13} by electronic fields. 
Topological responses such as the formation of a spin hedgehog \cite{Mil13, Sch14} and the topological Hall effect \cite{Jia17, Lit17} have also been extensively investigated.
These outstanding features have motivated the development of new information storage and logic technologies using magnetic skyrmions.
This research trend has also been stimulating the study of thin films of chiral materials instead of single crystals.

The CSL, which is a prototype of nonlinear chiral spin order \cite{Dzy58, Izy84, Kis05}, consists of forced ferromagnetic (FFM) regions uniformly separated by spin-twisting regions called chiral solitons. 
CSL formation was reported in chiral monoaxial crystals of CrNb$_3$S$_6$ \cite{Tog12} and YbNi$_3$Al$_9$ \cite{Aok18}, as well as in thin films of MnSi \cite{Wil13} and FeGe \cite{Kan16, McG16} under uniaxial strain.
In the CSL phase, the magnetoresistance (MR) exhibits a discrete change in resistance values, thus providing information on the number of chiral solitons confined in the sample \cite{Tog15}.
A discretized response was also found in the microwave regime in magnetic resonance experiments \cite{Gon17, Shi20}. 
Such a topological nature of the CSL is also favorable for promoting device applications such as multivalued memory.

Establishing a method for thin film growth of chiral magnets is inevitable for utilizing these striking characteristics of chiral spin order for device applications. 
Moreover, carrier density control is feasible in thin films by applying a strong electric field \cite{Uen08, Ye10}, which may enable us to control material parameters such as $T_{\rm m}$ and the critical field $H_{\rm c}$, as well as the soliton density, instead of using magnetic fields. 
Controlling soliton density by using electric fields in thin films may have device applications and may further promote their utilization.
Thin film growth thus is an important first step towards the implementation of chiral magnetic device applications.

In this study, we report the successful thin film growth of YbNi$_3$Al$_9$ using molecular beam epitaxy (MBE). 
YbNi$_3$Al$_9$ is a rare-earth-based monoaxial chiral magnet \cite{Oha14, Mat17}, which exhibits the CSL phase under magnetic fields \cite{Aok18}.
This material is the first example of the presence of the CSL phase in heavy fermion compounds, which involve strong electron-electron correlations. 
Thin films of multidomain YbNi$_3$Al$_9$ were obtained on $c$-plane sapphire substrates, in which the $c$-axis grew preferentially within the substrate planes.
The temperature dependence of electrical resistivity qualitatively reproduced the behavior of bulk crystals, including the dense Kondo effect and CHM formation.
The magnetic phase diagram is consistent with that previously obtained for bulk crystals, implying that the CSL state arises in the thin film samples.

\begin{figure}
\begin{center}
\includegraphics[width=6cm]{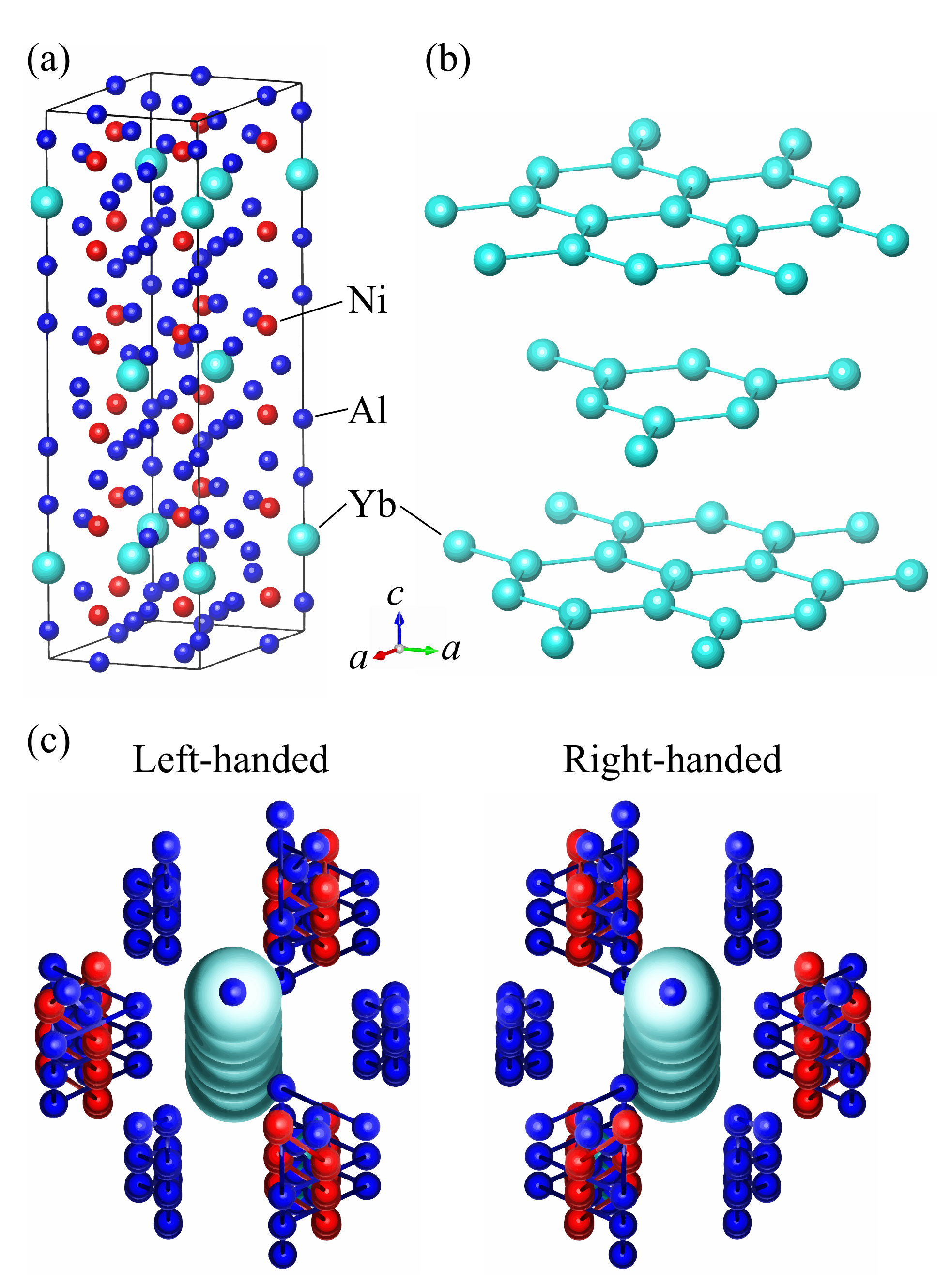}
\end{center}
\caption{\label{crystal}(a) Crystal structure of YbNi$_3$Al$_9$ and (b) honeycomb planes of Yb atoms. (c) Ni and Al atom alignment around the Yb atom in left- and right-handed crystals.}
\end{figure}

YbNi$_3$Al$_9$ crystallizes in a trigonal ErNi$_3$Al$_9$-type crystal structure with lattice constants of $a$ = 7.2731\,\AA\ and $c$ = 27.364\,\AA, as shown in Fig.~\ref{crystal}(a) \cite{Oha11, Yam12, Nak20}.
The space group of this compound is \#155 $R$32, which belongs to the Sohncke space group.
Yb atoms, which occupy 6$c$ Wyckoff sites, are arranged in honeycomb planes and are stacked sequentially, as shown in Fig.~\ref{crystal}(b).
Left- or right-handed helical configurations are formed by Ni (18$f$ site) and Al atoms (9$d$, 9$e$, and 18$f$ sites) along the principal $c$-axis of the three-fold screw symmetry, as shown in Fig.~\ref{crystal}(c). 
The distance between the nearest neighbor Yb atoms is 4.199\,\AA, while that between the adjacent planes is 9.121\,\AA.
Such a large difference implies a two-dimensional magnetic nature for this compound.
In fact, the critical exponents of magnetization were found to be close to those in the two-dimensional Ising model \cite{Wan20}. 
The valence of Yb ions was reported as 2.95 at 22\,K by hard X-ray photoemission spectroscopy \cite{Uts12}, ensuring that Yb atoms have a nonzero total angular momentum. 
The magnetic susceptibility was well fitted by the Curie--Weiss law above a temperature of 50\,K, with an effective magnetic moment of $\sim$4.4$\mu_{\rm B}$ \cite{Oha11, Yam12}.
This value is consistent with those expected for Yb atoms that have a valence near 3+. 
The magnetic part of the electrical resistivity of YbNi$_3$Al$_9$ exhibits a $-\log T$ dependence and has a maximum at $\sim$40\,K.
Below 40\,K, the resistivity decreases with decreasing temperature \cite{Oha11, Yam12}.
These features are typical characteristics of dense Kondo systems. 
The electronic specific heat coefficient reaches 110\,mJ/K$^2\cdot$ mol, indicating the heavy fermion character of this compound \cite{Yam12, Miy12}.

The CHM order sets in along the $c$-axis with the propagation vector $q$ $\simeq$ (0, 0, 0.82) \cite{Mat17} below $T_{\rm m}$ = 3.4\,K \cite{Oha11, Yam12, Miy12} in YbNi$_3$Al$_9$. 
The CHM state transforms into the FFM state at $H_{\rm c}$ = 1\,kOe when a magnetic field is applied in a direction perpendicular to the $c$-axis \cite{Yam12, Miy12}. 
Discrete changes in the MR were observed below $H_{\rm c}$ in the same configuration of magnetic fields in micrometer-sized samples \cite{Aok18}. 
This result indicates that the CSL state is realized in this compound.
The CSL formation is also supported by the magnetic entropy changes \cite{Wan20_2}.

The CSL phase region is highly enlarged in the field and temperature phase diagrams when substituting Ni with Cu, as expressed by Yb(Ni$_{1-x}$Cu$_x)_3$Al$_9$. 
For instance, $T_{\rm m}$ increases to 6.4\,K and $H_{\rm c}$ reaches up to 10\,kOe in $x = 0.06$ samples \cite{Oha14, Mat17, Nin18}. 
It must be noted that in the Cu-substituted system, the existence of the CSL state was confirmed as a development of the peaks of higher harmonics when using resonant X-ray diffraction (XRD) \cite{Mat17}.

The MBE technique was employed for the thin film growth of YbNi$_3$Al$_9$ in this study. 
We optimized the substrate temperature and atomic deposition rate ratio to obtain good crystallinity of the films.
After the film growth, the crystalline structure of the grown films was examined using XRD, while the top surface flatness was studied by atomic force microscopy (AFM). 
The resistivity measurements were performed using a standard four-probe method in a physical property measurement system (PPMS, Quantum Design).

We found that YbNi$_3$Al$_9$ was grown on $c$-plane sapphire substrates at 750\,$^\circ$C. 
Sapphire substrates were thermally treated at 1000\,$^\circ$C for 100 minutes at ultra-high vacuum prior to deposition to improve the surface flatness by promoting rearrangement of surface atoms. 
The pressure during the deposition was $\sim$$2 \times 10^{-6}$\,Pa. 
Elementary materials, namely, Yb, Ni, and Al were deposited using Knudsen cells with a stoichiometric atomic deposition rate ratio of 1:3:9 for Yb, Ni, and Al, respectively. 
The deposition rate of the film was maintained at 0.21\,\AA/s to achieve a thickness of 200\,nm.



\begin{figure}
\begin{center}
\includegraphics[width=13cm]{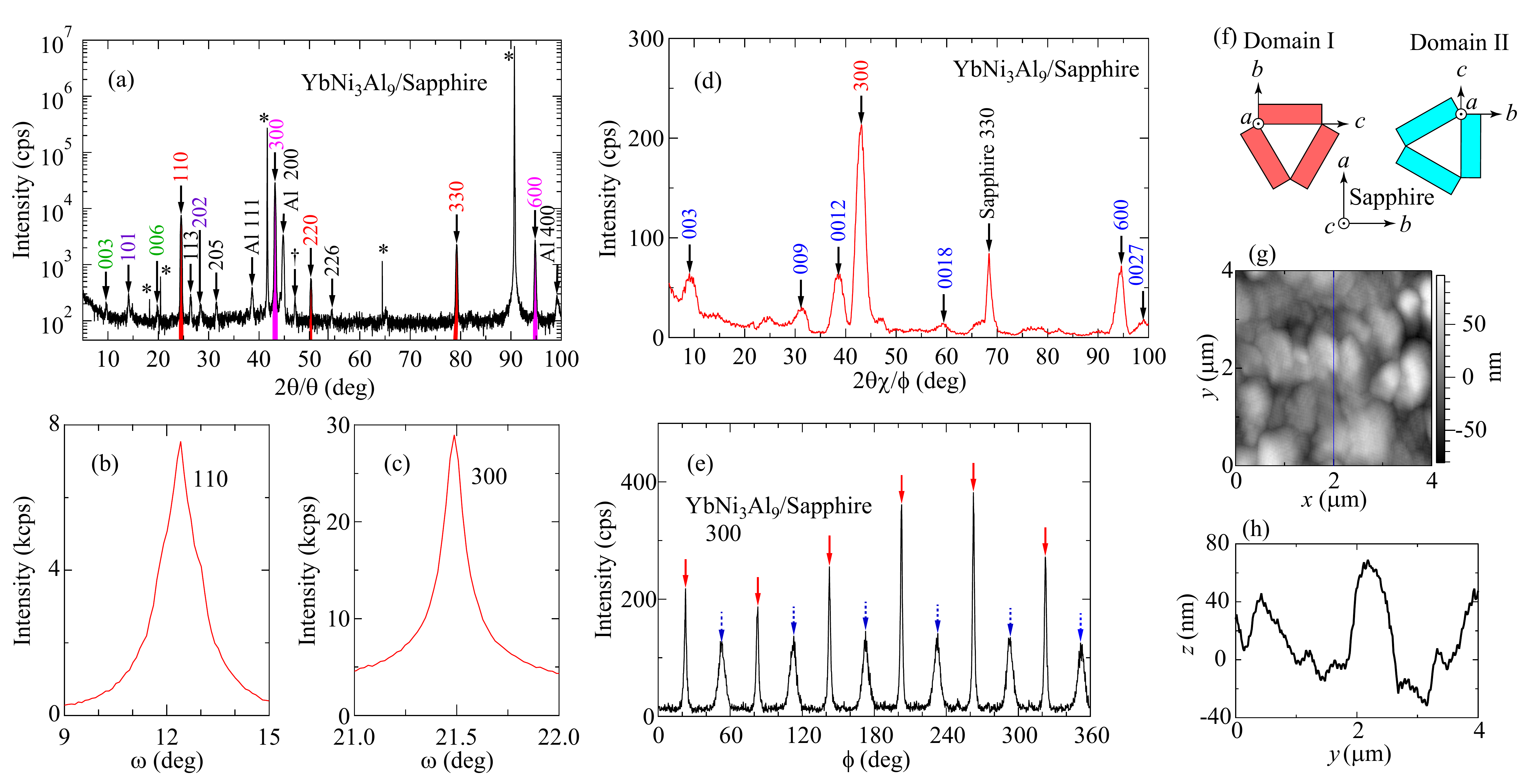}
\end{center}
\caption{\label{XRD}(a) XRD pattern in the direction perpendicular to the film surface for YbNi$_3$Al$_9$ on the $c$-plane sapphire substrate. Substrate peaks are marked by asterisks. The unknown impurity peak is marked by a dagger. Rocking curves for (b) 110 and (c) 300 peaks. (d) In-plane XRD pattern. (e) $\phi$-scan of the YbNi$_3$Al$_9$ 300 peak. Two different orientation domains result in two groups of six peaks, shown by red solid arrows and blue dotted arrows. (f) Schematic illustrations of two different orientation domains of YbNi$_3$Al$_9$ on the $c$-plane sapphire. (g) AFM image. (h) Line profile along the solid line in (g).} 
\end{figure}

Figure~\ref{XRD}(a) shows typical XRD data for the films grown on the $c$- plane of the trigonal sapphire substrate. 
The main peaks, highlighted in color, correspond to the $h$$h$0 and 3$h$00 peaks of YbNi$_3$Al$_9$.
Concomitantly, weak 00$l$ and $h$0$h$ peaks associated with YbNi$_3$Al$_9$ were observed. 
These observations indicate that domains oriented along the $a$ ([100]) and $b$ ([120]) axes are predominantly grown on the substrate, while a small portion of $c$-axis- and [211]-oriented domains develop in the film. 
Other weak peaks were derived from polycrystalline and oriented Al components.
An unidentified peak at $\sim$47$^\circ$ is probably due to an impurity.

The quality of the crystalline orientation of the grown films was examined by the XRD rocking-curve measurements. 
The film presented in Fig.~\ref{XRD} has 0.73$^\circ$ full width at half maximum (FWHM) for the 110 peak and 0.19$^\circ$ FWHM for the 300 peak, as shown in Figs.~\ref{XRD}(b) and \ref{XRD}(c), respectively. 
These data indicate that the YbNi$_3$Al$_9$ films have relatively high orientation and contrast with the orientation quality previously obtained in thin films of heavy fermion systems \cite{Hut93}.

The predominant orientation of the film was also evaluated by the grazing-incidence XRD profile, as shown in Fig.~\ref{XRD}(d).
Here, the X-ray incident direction was parallel to the $a$-axis of the sapphire substrate.
The 300 and 00$l$ peaks of YbNi$_3$Al$_9$ were observed along with the sapphire 330 peak, indicating the presence of in-plane-oriented domains for thin films. 
It directly signifies the existence of epitaxial domains. 
Here, we note that the $b$- and $c$-axes are perpendicular to the $a$-axis.
The $a$-axis-oriented domain, as discussed for Fig.~\ref{XRD}(a), grows epitaxially, while the $b$-axis-oriented domain has no in-plane orientation. 
Figure~\ref{XRD}(e) shows the $\phi$-scan pattern of the 300 YbNi$_3$Al$_9$ peak.
Six sharp peaks indicated by red solid arrows appear every 60$^\circ$, reflecting the six-fold symmetry of the substrate. 
Concomitantly, six relatively broad peaks indicated by blue dotted arrows are seen every 60$^\circ$.

Based on a comprehensive understanding of the XRD data, we can conclude the following characteristics of the grown films:
(1) Three equivalent epitaxial grains (domain I) were grown with YbNi$_3$Al$_9$ $b$-, $c$-, and $a$-axes parallel to the sapphire $a$-, $b$-, and $c$-axes, as schematically illustrated in Fig.~\ref{XRD}(f). 
Domain I corresponds to $h$$h$0 and to the 300 peaks in Figs.~\ref{XRD}(a) and \ref{XRD}(d), respectively.
(2) Domain II is also obtained, the relative orientation of which is rotated by 30$^\circ$ in the $\phi$ direction from that of domain I.
The three equivalent epitaxial grains exist with YbNi$_3$Al$_9$ $c$-, $b$-, and $a$-axes parallel to the sapphire $a$-, $b$-, and $c$-axes, as shown in Fig.~\ref{XRD}(f).
Domain II corresponds to the $h$$h$0 and 00$l$ peaks in Figs.~\ref{XRD}(a) and \ref{XRD}(d), respectively.
(3) The $\phi$-scan data suggest that domain I is highly oriented in comparison with domain II. 
(4) A small portion of the film consists of misaligned domains. 
They may have $b$-axis, $c$-axis, and [211] orientations perpendicular to the substrate surface. 
Alternatively, it is likely to have randomly distributed orientations and thus form polycrystalline structures.

The lattice constants of the epitaxial grains were obtained from the interplane and in-plane XRD data. 
The lattice constant along the $a$-axis, which is oriented in the direction normal to the substrate surface, is estimated to be 7.258(4)\,\AA\ and is in good agreement with the reported value \cite{Oha11, Yam12} of 7.2731\,\AA, being within $\sim$ 0.2\%. 
The lattice constants along the $b$- and $c$-axes parallel to the substrate surface are estimated to be 6.23(1)\,\AA\ and 27.70(2)\,\AA, respectively. 
The deviations in the magnitude are of the order of $\sim$1\% from those of the bulk samples.
More accurately, the lattice shrinks along the $b$-axis, while it expands along the $c$-axis. 
These data indicate that the films are subject to in-plane lattice distortion. 

Sapphire has a trigonal corundum structure with lattice constants of $a = 4.759$\,\AA\ and $c = 12.99$\,\AA.
There are many possibilities of lattice matching between YbNi$_3$Al$_9$ and sapphire. 
For instance, a supercell structure consisting of two unit cells of YbNi$_3$Al$_9$ along the $a$-axis and three unit cells of sapphire along the $a$-axis provides a lattice mismatch of 1.7\%.
Different supercell structures made of 3 unit cells of YbNi$_3$Al$_9$ along the $c$-axis and 17 units cells of sapphire along the $a$-axis yield a lattice mismatch of 1.3\%, which is slightly smaller than the value in the former case.
Another combination with the $b$-axis of sapphire exhibits smaller values of lattice matching: 0.8\% for the $c$-axis of YbNi$_3$Al$_9$ by matching 3 unit cells of YbNi$_3$Al$_9$ and 20 unit cells of sapphire. 
Almost the same value of 0.8\% is obtained with the $b$-axis of YbNi$_3$Al$_9$ when two unit cells of YbNi$_3$Al$_9$ match three unit cells of sapphire. 
These possible combinations may induce the growth of multigrain structures in the YbNi$_3$Al$_9$ films.

An island-like morphology on the surface with a typical size of 0.5--1 $\mu$m in diameter was confirmed in the AFM image and the line profile, as shown in Figs.~\ref{XRD}(g) and \ref{XRD}(h), respectively.
The island height ranges from $\sim$40 to $\sim$120\,nm. 
In addition, protuberances with a diameter of 2--3 $\mu$m were observed (not shown). 
Excess Al may give rise to such debris on the surface during the growth process.

\begin{figure}
\begin{center}
\includegraphics[width=7cm]{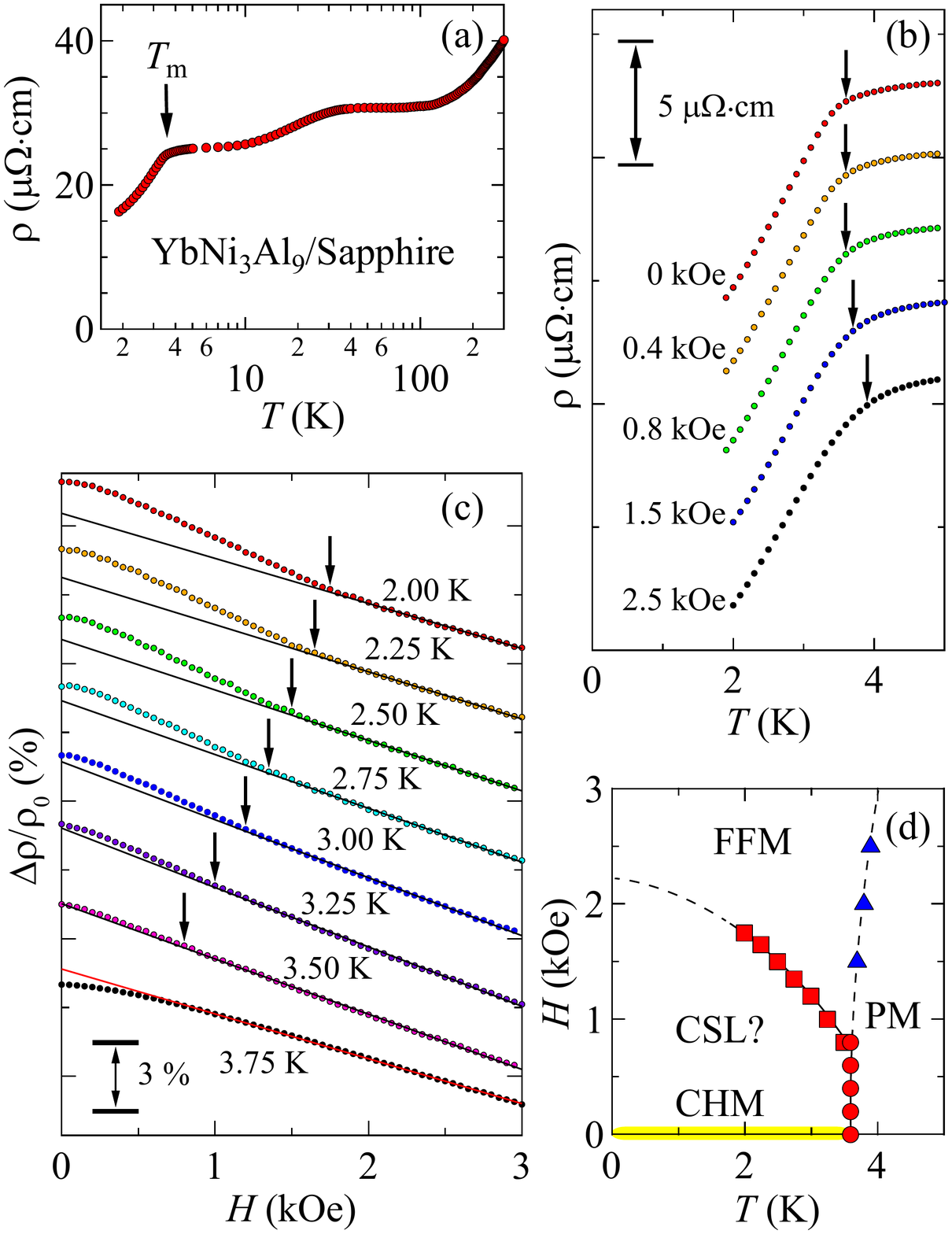}
\end{center}
\caption{\label{resis}(a) Temperature dependence of the resistivity and (b) that for low temperatures under various magnetic fields. (c) Magnetoresistance at various temperatures. Each curve is shifted for clarity. Solid lines indicate linear fit regression curves in the forced ferromagnetic region. (d) Magnetic phase diagram obtained from the temperature dependence of the resistivity (\scriptsize$\bigcirc$ and $\triangle$) and magnetoresistance ($\square$). Solid and dotted lines are only a guide for readability.}
\end{figure}

Figure~\ref{resis}(a) shows the temperature dependence of the electrical resistivity. 
The resistivity decreases with decreasing temperature, starting from room temperature.
It exhibits a broad shoulder structure at $\sim$40\,K, corresponding to the Kondo effect. 
Here, we note that the resistivity shown in Fig.~\ref{resis}(a) includes both magnetic and phonon contributions, obscuring the -$\log T$ dependence.
Below 40\,K, the resistivity decreases smoothly by constructing a coherent electrical state \cite{Hew93}.
It exhibits a kink structure at $T_m = 3.6$\,K, as indicated by the arrow.
The overall temperature dependence qualitatively reproduces that for bulk crystals; therefore, we probably conclude that the CHM order sets in below $T_m$ even in the thin films, while the transition temperature is slightly higher than that in bulk crystals.

The temperature dependence of resistivity was measured at various magnetic fields to examine the existence of chiral spin order, as shown in Fig.~\ref{resis}(b).
Here, the fields were applied in the direction perpendicular to the substrate plane, namely, perpendicular to the $c$-axis for major domains of YbNi$_3$Al$_9$ thin films. 
The kinks appear at almost the same temperature below 0.8\,kOe, and they tend to shift toward higher temperatures with further increasing fields. 
Under the assumption that the CSL state arises in thin films, the former feature corresponds to the CSL phase transition from the paramagnetic (PM) phase. 
The latter indicates a crossover from PM to FFM states.

MR data measured at low temperatures are shown in Fig.~\ref{resis}(c).
The distance between the voltage terminals was 2.75\,mm, which was much longer than the CHM period.
We focused on the observation of the phase boundary between the CSL and the FFM region.
Negative MR behavior was observed in the thin films, while the magnitude of the MR was largely suppressed (by 30\%) in comparison with that for bulk single crystals in the FFM state\cite{Miy12}. 
This may be caused by the sample quality of the films, including the presence of in-plane multidomain structures.
The slope of the MR exhibits a finite change at a particular magnetic field, as indicated by arrows, the value of which corresponds to $H_{\rm c}$, as discussed later.
The MR below $H_{\rm c}$ is higher than the linear fit regression curves extrapolated from the FFM region, as shown by the solid lines. 
This feature is no longer observed above $T_{\rm m}$ and is qualitatively consistent with that obtained in bulk crystals. 
Based on previous studies, the most plausible origin is ascribed to the phase transition from the CSL phase to the FFM state. 

The magnetic phase diagram was obtained from the temperature and field dependence of resistivity, as summarized in Fig.~\ref{resis}(d). 
A crossover line between the PM phase and FFM state merges with another phase transition line at $\sim$0.8\,kOe.
This phase transition line identified by the MR anomaly corresponds to $H_{\rm c}$ for the CSL formation. 
The $H_{\rm c}$ line increases with decreasing temperature, reaching 1.75\,kOe at the lowest measuring temperature of 2\,K in the experiments. 
All the observed features qualitatively reproduced the magnetic phase diagram previously reported in bulk crystals, while the $H_{\rm c}$ line shifts toward higher fields owing to a strong diamagnetic effect in thin films. 
A similar enhancement of $H_{\rm c}$ was also reported in micrometer-sized samples \cite{Aok18}, which were also subject to the same effect.
This agreement implies that the CSL state arises even in thin film samples.

Direct observation of the CSL state was achieved by transmission electron microscopy (TEM) for CrNb$_3$S$_6$ \cite{Tog12}. 
However, it cannot be applied to YbNi$_3$Al$_9$ because of the extremely low $T_{\rm m}$ even when using a cutting-edge cryogenic TEM machine \cite{Tog21}.
Another useful experimental method involves polarized neutron reflectometry \cite{Mey17} and resonant elastic X-ray scattering \cite{Zha17}. 
Not only extremely low $T_{\rm m}$ values, but also the rough surface of the present films may render it difficult to perform these measurements because the neutron beam or X-rays irradiate the thin films at small incident angles. 
Direct observation of the CSL state in the present films is a challenge for the future.
 

In summary, we have achieved in growing thin films of the heavy fermion chiral magnet YbNi$_3$Al$_9$ using MBE. 
The establishment of a thin film growth method for rare-earth-based chiral magnets may provide useful information on the chiral spin system and pave way for device applications such as multivalued memory using the topological CSL state.
In particular, carrier density control with the use of electric fields may be available in thin films.
This may provide a promising method to tune various physical properties of thin film devices.
It may change the Kondo temperature, Ruderman--Kittel--Kasuya--Yosida interaction, Dzyaloshinskii--Moriya interaction strength, and $H_{\rm c}$ and $T_{\rm m}$ values for the CHM and CSL phases.  
We believe our study will contribute to advancing our understanding of chiral magnetism from the viewpoints of both fundamental physics and device applications.

\begin{acknowledgments}
This work was supported by Grants-in-Aid for Scientific Research (Nos. 19K03751, 17H02767, and 18K03539) and by an Innovative Areas ``Quantum Liquid Crystals'' grant (KAKENHI Grant No. JP19H05826) from the Japan Society for the Promotion of Science.
\end{acknowledgments}

\section*{DATA AVAILABILITY}
The data that support the findings of this study are available from the corresponding author upon reasonable request.

\end{document}